\documentclass[twocolumn,nofootinbib,superscriptaddress]{revtex4-1}

\pdfoutput=1

\usepackage{slashed}

\usepackage[normalem]{ulem}

\usepackage{color}

\usepackage{epsfig}
\usepackage{epstopdf}
\usepackage{epsf}
\input{epsf.sty}
\usepackage{graphicx,amsmath,amssymb}
\usepackage{epsfig}
\allowdisplaybreaks

\def\ei{\end{itemize}}
\def\be{\begin{equation}}
\def\ee{\end{equation}}
\newcommand{\bea}{\begin{eqnarray}}
\newcommand{\eea}{\end{eqnarray}}

\usepackage{bbm,bm,amsmath,amssymb}

\usepackage{hyperref}

\begin{document}

\title{Observable Chiral Gravitational Waves from Inflation in String Theory}

\author{Evan McDonough}
\email{evan\_mcdonough@brown.edu}
\author{Stephon Alexander}
\email{stephon\_alexander@brown.edu}
\affiliation{Department of Physics, Brown University, Providence, RI, USA. 02903}

 \begin{abstract}
We consider gravitational wave production during inflation in type IIB string theory, and the possibility of observable gravitational waves in small field inflation. We show that the gauge field excitations on a set of coincident D7 branes, itself critical for moduli stabilization and hence intrinsic to inflation in string theory, coupled with axion fields from bulk fluxes, can act as a spectator sector during inflation. This results in a large production of chiral gravitational waves, even for relatively small values of the axion-gauge field coupling. We extend this to include a monodromy for the axion, and demonstrate that in both cases an observable level of gravitational waves is produced in small field inflation in string theory, with a spectrum that is maximally chiral. Finally, we demonstrate the consistency with moduli stabilization and with arbitrary (large or small field) inflationary dynamics of the host model, considering as an explicit example Kahler Moduli Inflation.
\end{abstract}

\maketitle

\section{Introduction}

Observable primordial gravitational waves from inflationary cosmology are typically associated with large field models, as required by the Lyth bound \cite{Lyth:1996im}. Unfortunately, these models generically suffer from the $\eta$-problem \cite{Copeland:1994vg}, motivating an analysis from a UV complete framework, such as string theory. However, even in string theory, large field inflation faces severe challenges, notably from back-reaction on string theory moduli \cite{Conlon:2011qp,Hebecker:2014kva,Baumann:2014nda} and the Weak Gravity Conjecture \cite{Hebecker:2015zss, *Ibanez:2015fcv, *Heidenreich:2015wga, *Brown:2015lia,*Brown:2015iha, *Rudelius:2015xta,*Rudelius:2014wla}.

Fortunately, intrinsic to string theory realizations of inflation is the existence of much more structure then simply an inflaton and its potential energy, e.g.~\cite{Alexander:2001ks,Burgess:2001fx,Dvali:2001fw} (for a comprehensive review see \cite{Baumann:2014nda}). All constructions require that the additional moduli, potentially numbering in the hundreds, be stabilized. We interpret this not as a drawback or complication, but a \emph{feature} of inflation in string theory, which may lead to distinct observational signatures.  

With this goal in mind, in this work we consider the production of gravitational waves from the dynamics of these additional fields, and find that an observable level of (chiral) gravitational waves can generically arise in both large and small field models of inflation in string theory.

More concretely, the stabilization of string theory moduli is most often achieved (e.g.~\cite{Kachru:2003aw,Balasubramanian:2005zx, Conlon:2005ki}) via a combination of fluxes and \emph{gaugino condensation}. The latter is captured by a simple superpotential,
\be
W = e^{ - \frac{2\pi}{N} \hat{f}_{D7}} ,
\ee
where $N$ is the rank of the $SU(N)$ gauge theory on worldvolume of $N$ coincident D7 branes, and $\hat{f}_{D7}$ is the gauge kinetic function. This generates exponential potentials for volume moduli and oscillatory potentials for axions, accompanied by couplings ${\rm Re}\hat{f}_{D7}F^2$ and ${\rm Im}\hat{f}_{D7} F \tilde{F}$ of the moduli to the $SU(N)$ gauge field. This complex system of interacting axions and non-Abelian gauge fields is present in all string inflation scenarios, and could in principle be evolving during inflation. Similarly, the possibility that a string theory axion could be slowly-rolling but not itself be the driver of inflation was first considered in \cite{Alexander:2004xd} (see also \cite{Kobayashi:2009cm}).

In \cite{Dimastrogiovanni:2016fuu} it was shown that the dynamics of precisely such a `spectator sector', i.e. a set of fields decoupled from the inflaton and subdominant in energy density, could lead to significant production of gravitational waves on cosmological scales. This construction evades the Lyth bound \cite{Lyth:1996im}, raising the interesting possibility that observable gravitational waves may not require large field inflation \cite{Fujita:2017jwq}. Moreover, this occurs without a significant production of scalar perturbations, in contrast with other mechanisms for gravitational wave production in string theory \cite{Mirbabayi:2014jqa}.

In this paper we realize variants of \cite{Dimastrogiovanni:2016fuu} via the interacting system of a $C_2$ axion and the $SU(N)$ gauge fields on a set of D7-branes. Worldvolume fluxes on an internal two-cycle generate a $C_2$ dependence of the gauge kinetic function, giving rise to an oscillatory potential for the axion and an axion-gauge field coupling that is enhanced by a factor of $N$. We extend this to include a monodromy potential for the axion \cite{McAllister:2008hb, Silverstein:2008sg}, and show these scenarios can be consistently realized in the large volume scenario (LVS) \cite{Balasubramanian:2005zx, Conlon:2005ki}, without backreaction, and independent of the details of inflation. 

We demonstrate that gravitational waves are copiously produced, though for a periodic axion potential the production generically (i.e. barring extremely large $N$) occurs only for the first few e-folds of inflation. With the inclusion of a monodromy potential, the gravitational wave production is long lasting, occurring for tens of e-folds.

Provided that the CMB pivot scale exited the horizon while gravitational waves are being produced, the result of both cases is an amplification of the tensor-to-scalar ratio, and a net \emph{chirality} on large scales. This leads to an tensor-to-scalar ratio of $r\simeq 10^{-3} - 10^{-2}$ in small field inflation, observable by next generation Cosmic Microwave Background B-mode polarization experiments \cite{Abazajian:2016yjj}, along with parity violating CMB cross correlations \cite{Lue:1998mq, Contaldi:2008yz}, of which the observational prospects have been studied in \cite{Gluscevic:2010vv,Gerbino:2016mqb}. This is particularly interesting given the role of chiral gravitational waves in models of leptogenesis  \cite{Alexander:2004us, Maleknejad:2014wsa,Maleknejad:2016dci, Adshead:2017znw, Caldwell:2017chz} and dark matter \cite{Alexander:2018fjp}.

Finally, as a concrete and complete model, we embed this into Kahler Moduli Inflation \cite{Conlon:2005jm}, with moduli stabilization occurring as in the LVS. This inflation model predicts $r\leq10^{-10}$, and we show that with the inclusion of the spectator sector, this can be raised to an observable level.

The structure of this paper is as follows: we first review the mechanism for production of gravitational waves put forward in \cite{Dimastrogiovanni:2016fuu}, and its generalization from $SU(2)$ to $SU(N)$. In Section III we discuss the realization of this via $C_2$ axions and gaugino condensation on branes carrying worldvolume flux, and in Section IV we consider the effect of a monodromy potential for the axion. In Section V we implement this in the context of the Large Volume Scenario, and demonstrate the consistency with moduli stabilization and arbitrary inflationary dynamics.  In section VI we consider an example realization in the context of Kahler Moduli Inflation. We close in section VII with a discussion of directions for future work.

\vspace{-3mm}
%=================================================================%
\section{Gravitational Waves from Axion-Gauge Field Dynamics}
\label{sec:GWs}
%=================================================================%

\subsection{The DFF Model: Spectator SU(2)}

In this work we study the production of gravitational waves in slight variants of the model \cite{Dimastrogiovanni:2016fuu}. This is a spectator version of the inflationary model \cite{Adshead:2012kp,Adshead:2013nka} (see also \cite{Alexander:2011hz}), which in turn builds upon \cite{Freese:1990rb,Adams:1992bn}. The models we study have the nice feature that while the inflation model \cite{Adshead:2012kp,Adshead:2013nka} is ruled out on observational grounds \cite{Adshead:2013nka}, the spectator models inherit the scalar perturbation predictions (e.g.~$n_s$) of their `host' model, which can be arbitrarily chosen.

This can be succinctly described the matter Lagrangian,
\be
\label{DFFL}
\mathcal{L} = \mathcal{L}_{\phi}[\phi] + \mathcal{L}_{GW}[\chi,F] ,
\ee
where  $\mathcal{L}_{\phi}$ is the inflaton and its potential, while $ \mathcal{L}_{GW}$ is the axion-gauge field system responsible for the production of gravitational waves. The field $\chi$ slowly rolls down its potential during inflation, and the fixed sign of $\dot{\chi}$ results in a gravitational wave spectrum that is maximally chiral.

In \cite{Caldwell:2017chz} it was shown that the chiral gravitational waves in a variant of this scenario can lead to successful baryogenesis, and in \cite{Alexander:2018fjp} that model was used to construct a new class of dark matter models, dubbed the Dark Baryon Superfluid. The observational predictions of \cite{Dimastrogiovanni:2016fuu} were further refined in \cite{Agrawal:2017awz, Thorne:2017jft, Dimastrogiovanni:2018xnn}, where it was shown that a large tensor non-Gaussianity can be produced in these models, without an accompanying production of scalar non-Gaussianity (see also Appendix \ref{App}). Similar phenomena can also arise in the case that the gauge field is Abelian, given suitable values of the model parameters \cite{Namba:2015gja}.

The Lagrangian for the gravitational wave sector in \cite{Dimastrogiovanni:2016fuu} is given by,
\be
\label{LGW}
\mathcal{L}_{GW} = -\frac{1}{2}\left(\partial\chi\right)^{2}-U(\chi)-\frac{1}{4}F_{\mu\nu}^{a}F^{a\mu\nu}+\frac{\lambda\,\chi}{4f}F_{\mu\nu}^{a}\tilde{F}^{a\mu\nu} ,
\ee
where $F_{\mu\nu}^{a}$ is the non-Abelian field strength with coupling $g$. During inflation the dynamics of the universe are predominantly determined by the inflaton $\phi$, with the axion and gauge field vastly subdominant in energy density. Moreover, as in  \cite{Adshead:2012kp,Adshead:2013nka}, the gauge field $A_\mu$ is chosen to take the isotropic configuration,
\begin{equation}
\langle A_{0}^{a} \rangle=0\;\;,\;\; \langle A_{i}^{a} \rangle=\delta_{i}^{a} a(t) Q(t)\,.
\end{equation}
The equations of motion for the axion-gauge field system are then given by,
\begin{eqnarray}
&& \ddot{\chi}+3H\dot{\chi}+U_{,\chi}=-\frac{3g\lambda}{f}Q^2 \left(\dot{Q}+HQ\right)  \label{EOMchi}\,,\\
&& \ddot{Q}+3H\dot{Q}+\left(\dot{H}+2H^2\right)Q+2g^2 Q^3=\frac{g\lambda}{f}\dot{\chi}Q^2 \label{EOMQ}   \,.
\label{gauge-bck}
\end{eqnarray}
The potential for $\chi$ and initial conditions for $\chi$ and $Q$ are chosen that they are both slowly-rolling during inflation. There is an attractor solution wherein $Q$ and $\chi$ non-trivially balance one another, which occurs provided
\be
\label{cond1}
\frac{\lambda Q}{f} \gg 1 ,
\ee
and
\be
\label{cond2}
 \frac{\lambda Q}{f} \cdot  \frac{g Q}{H}  \gg 1 .
\ee
This imposes constraints on the parameters of the model, as we will come back to shortly.

The slowly rolling attractor solution is given by \cite{Dimastrogiovanni:2012ew},
\be
Q  \simeq  \left[ \frac{- f}{3 g\lambda H} U_{,\chi} \right]^{1/3} \;\; ,\;\; \label{EOMchiSR}
\dot{\chi} \simeq  \frac{2 f g}{\lambda}  Q  .
\ee
These equations accurately describe the evolution of the spectator axion-gauge field system during inflation provided that the parameters $\{ g,\lambda,f,\mu\}$ are such that the conditions \eqref{cond1} \eqref{cond2} are satisfied.

\subsection{Gravitational Waves}

The color-spatial fluctuations in the gauge field $\delta A_{\mu a}$ can be decomposed into two scalars (e.g.~$\delta Q \delta_{ia}$), two vectors (e.g.~$\partial_i M_a$), and two tensor polarizations ($\delta A_{ia} = \delta_a \, ^j t_{i j}$). The tensor fluctuations $t_{ij}$ are amplified due to the dynamics of $\chi$, and this sources gravitational waves at linear order in perturbations, via the combination $\langle A \rangle \delta A \sim Q t_{ij}$. For a detailed discussion see e.g.~\cite{Adshead:2013nka, Dimastrogiovanni:2012ew, Caldwell:2017chz, Maleknejad:2016qjz}.

 The equations of motion for the coupled metric and gauge field tensor fluctuations are given by \cite{Dimastrogiovanni:2012ew}
\bea
&& \partial_x ^2 t_{R,L} + \left[ 1 + \frac{2 m_Q \xi}{x^2} \mp \frac{m_Q + \xi}{x}\right]t_{R,L} \simeq0 , \\
&& \label{eqpsi} \partial_x ^2 \psi_{R,L} + \left( 1 - \frac{2}{x^2}\right)\psi_{R,L} \simeq \cal{S}^{\psi} _{RL} ,
\eea
where $x\equiv k/aH$, $\psi_{R,L}\equiv a M_{Pl} h_{ij}/2$ is the canonically normalized gravitational wave mode function, and $t_{R,L}\equiv a M_{Pl} t_{ij}/2$ is the gauge field tensor mode function. The parameters $\xi$ and $m_Q$ are defined as
\be
\xi \equiv \frac{\lambda \dot{\chi}}{2 f H} \;\; , \;\; m_Q \equiv \frac{g Q}{H} ,
\ee
which are positive during the slow-roll phase of $\chi$. This leads to an exponential growth of a single handedness of $t$, with solution \cite{Dimastrogiovanni:2012ew},
\be
t_{R} = \frac{1}{\sqrt{2k}} e^{\frac{\pi}{2} (m_Q + \xi)} W_{\alpha,\beta}\left( - \frac{2 i k}{a H} \right) .
\ee
There is no corresponding growth of scalar fluctuations provided that $m_Q > \sqrt{2}$ \cite{Dimastrogiovanni:2016fuu}.

This growth of $t_R$ is converted to a growth of $\psi_R$ via the source term on the right-hand-side of \eqref{eqpsi},
\be
\mathcal{S}^{\psi} _{R,L} \equiv (2  \sqrt{\epsilon_E}/x) \partial_x t_{R,L} +  (2  \sqrt{\epsilon_B}/x^2) (m_Q \mp x) t_{R,L} ,
\ee
where $\epsilon_E$ and $\epsilon_B$ are the slow-roll parameters of the electric and magnetic energy densities in the gauge field,
\be
\epsilon_E \equiv \frac{(\dot{Q}^2 + H Q)^2 }{H^2 M_{Pl}^2} \; ,\; \epsilon_B \equiv \frac{g^2 Q^4}{H^2 M_{Pl}^2} .
 \ee 
After solving for the $\psi$ using the Green's function method, the resulting sourced tensor power spectrum is given by
\be
P_h ^{s} (k) \simeq \frac{2}{\pi^2} \frac{H^2}{M_{Pl}^2} \cdot \epsilon_B e^{3.62 m_Q} |_{k=aH},
\ee
where $m_Q$ and $\epsilon_B$ are evaluated at $k=aH$. The tensor-to-scalar ratio is then given by a sum of the single-field and sourced tensor fluctuations,
\be
r \simeq r_{infl} \cdot \left(1 + \epsilon_B e^{3.6 m_Q}\right)|_{k= k_*} ,
\ee
where $r_{infl}$ is the single-field inflation prediction for $r$ and $k_*$ is the CMB pivot scale.

\subsection{Generalization to  $SU(N)$}

The gauge group of interest in the present work is not simply $SU(2)$, but $SU(N)$, where $N$ is the number of coincident D7-branes. In the original moduli stabilization proposal of KKLT  \cite{Kachru:2003aw}, gaugino condensation occurs on a stack of $N=20 \pi$ D7-branes, and we will consider similar values of $N$ here.

This introduces two important effects: (1) a rescaling of the string theory coupling $\lambda$, which we will see in detail in Section \ref{sec:C2}, and (2) the possibility of multiple excited independent subgroups of $SU(N)$.  In this subsection we focus on the latter, proceeding along the lines of Appendix A of \cite{Caldwell:2017chz} and  \cite{Caldwell:2018feo}. 

The production of gravitational waves in \cite{Dimastrogiovanni:2016fuu} is intrinsically dependent on the group $SU(2)$, and hence one must first split the $SU(N)$ into disjoint $SU(2)$ subgroups, of which the maximum number is $N/2$ mod 2.  
 
 While in this work we consider only a single $SU(2)$ subgroup to be excited, in principle one could consider a set of ${\cal N}$ $SU(2)$ subgroups with non-vanishing field strengths. The equations of motion take a simple form if the $SU(2)$ subgroups have a common field strength (which can be achieved e.g.~by initializing with a common initial condition), in which case the background equations of motion are equivalent to the single $SU(2)$ case \eqref{EOMchi} \eqref{EOMQ} with the replacement \cite{Caldwell:2017chz}
\be
g \rightarrow \sqrt{{\cal N}} g \;\; , \;\; Q \rightarrow \frac{Q}{\sqrt{{\cal N}}} .
\ee
Under this replacement, and ignoring the string theoretic rescaling of $\lambda$, the value of $m_Q$ is unmodified while $\epsilon_B \rightarrow \epsilon_B/{\cal N}$. On the other hand, the gravitational wave equation of motion now has ${\cal N}$ statistically uncorrelated sources.  The net effect is that the tensor to scalar ratio is \emph{unchanged} from the single $SU(2)$ case,
\be
\label{eq:r}
r \rightarrow r_{infl} \cdot \left( 1 + {\cal N}  \cdot \frac{\epsilon_B}{{\cal N}} e^{3.6 m_Q} \right) =  r_{infl} \cdot \left(1 + \epsilon_B e^{3.6 m_Q}\right) .
\ee
With this in mind, in this work we consider for simplicity the case that a single $SU(2)$ subgroup is excited, that is, we take ${\cal N} = 1$. 

We now consider the values of the other parameters describing the model.

\subsection{Constraints on Model Parameters for a Consistent Realization}

In order for the phenomena studied here to arise generically, there should exist a large region of parameter space that (i) realizes a self-consistent cosmological evolution of the axion and gauge field as a spectator sector, (ii) gives phenomenologically interesting results, and (iii) is self consistent from the perspective of the underlying string compactification.

To satisfy (i) we require that the slow-roll attractor exists, which requires \eqref{cond1} and \eqref{cond2} be satisfied. The first of these can re-phrased as a constraint on $f/H$ as a function of $\lambda$ and $m_Q$:
\be
\label{eq:f}
\frac{f}{H} \ll \frac{\lambda m_Q}{g} .
\ee
Interestingly, the slow-roll solution does not necessarily require $\lambda \gg 1$.

In order to have a sizeable amplification of the gravitational wave power spectrum, and hence satisfy (ii), it is required that
\be
m_Q > 1 ,
\ee
which in conjunction with \eqref{eq:f} implies that \eqref{cond2} is satisfied.

On the other hand, the time duration of the amplification can be approximated as the duration of slow-roll evolution. The slow-roll solution for $\chi$ can be written as,
\be
\frac{1}{f }\frac{{\rm d }{\chi} }{d N} = \frac{2 m_Q }{\lambda} .
\ee
In the absence of a monodromy potential, the amount of slow-roll evolution of $\chi$ is bounded by $\Delta \chi \simeq \pi f $, and in this case the number of e-folds of production is given by,
\be
\label{NGW}
N_{GW} \lesssim \frac{\pi \lambda}{2 m_Q} .
\ee
Hence, efficient gravitational wave production $(m_Q >1)$ will occur for more then a single e-fold for only for $\lambda >1$. This is modified in the presence of a monodromy potential, as we study in section \ref{sec:mono}. 

Regarding (iii), string theory examples generically lead to small values of the axion-gauge field coupling $\lambda$  \cite{Baumann:2014nda}. This is an obstacle to driving inflation using solely this configuration, as the number of e-folds of inflation is given in that case by \cite{Adshead:2012kp}
\be
N_{infl} \lesssim \frac{3}{5} \lambda ,
\ee
independent of $f$. In contrast, the spectator axion we study here need not slowly-roll for the full 60 e-folds of inflation, and thus the model can accommodate much smaller values of $\lambda$. We will see this in more detail in sections II-VI.

 As an example, consider tuning $\mu$ to make $m_{Q}=5$, and fix $\lambda=20$.  In this case, the production lasts for $N_{GW} \simeq 4$ e-folds of inflation. If in addition $g = 10^{-2}$ and $H_{inf}=10^{-7} M_{Pl}$ the constraint on $f$ reads: \be
f \ll  10^{-3} M_{Pl} ,
\ee
which allows for a large range of $f$. More generally, the model building challenge for these models, assuming a small $\lambda$, is to engineer a small enough $f$ to allow for slow roll dynamics. This is in stark contrast to the model building challenge of Natural Inflation  \cite{Freese:1990rb,Adams:1992bn}, where one must engineer a super-Planckian decay constant (for example, by alignment \cite{Kim:2004rp,Choi:2014rja}), in conflict with the Weak Gravity Conjecture \cite{Hebecker:2015zss, *Ibanez:2015fcv, *Heidenreich:2015wga, *Brown:2015lia,*Brown:2015iha, *Rudelius:2015xta,*Rudelius:2014wla}.

From this we conclude that there is a large parameter space of consistent and phenomenologically interesting models of an axion-gauge field spectator sector, and we will now proceed to study their realization in string theory.

%=================================================================%
\section{$C_2$ axions and Gaugino Condensation on Magnetized Branes}
\label{sec:C2}
%=================================================================%

The $C_2$ axion  was considered as a candidate for natural inflation in \cite{Long:2014dta} and \cite{Ben-Dayan:2014lca}. Here we use these constructions to build the $C_2$ axion as a spectator field, whose interactions with the gauge fields on a stack of D7 branes leads to an amplification of gravitational waves on large scales.

\subsection{Salient Points}

\begin{figure*}[ht!]
\centering
\includegraphics[width=8.5cm]{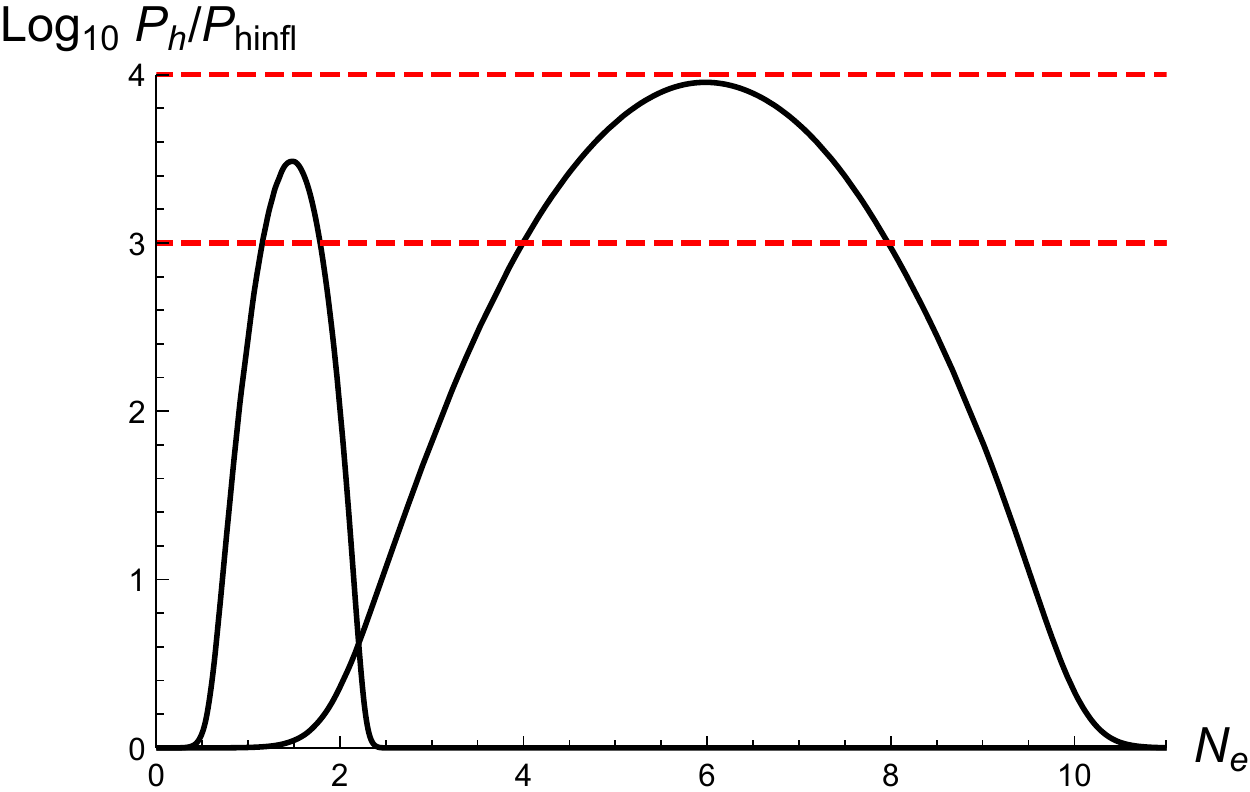}
\includegraphics[width=8.5cm]{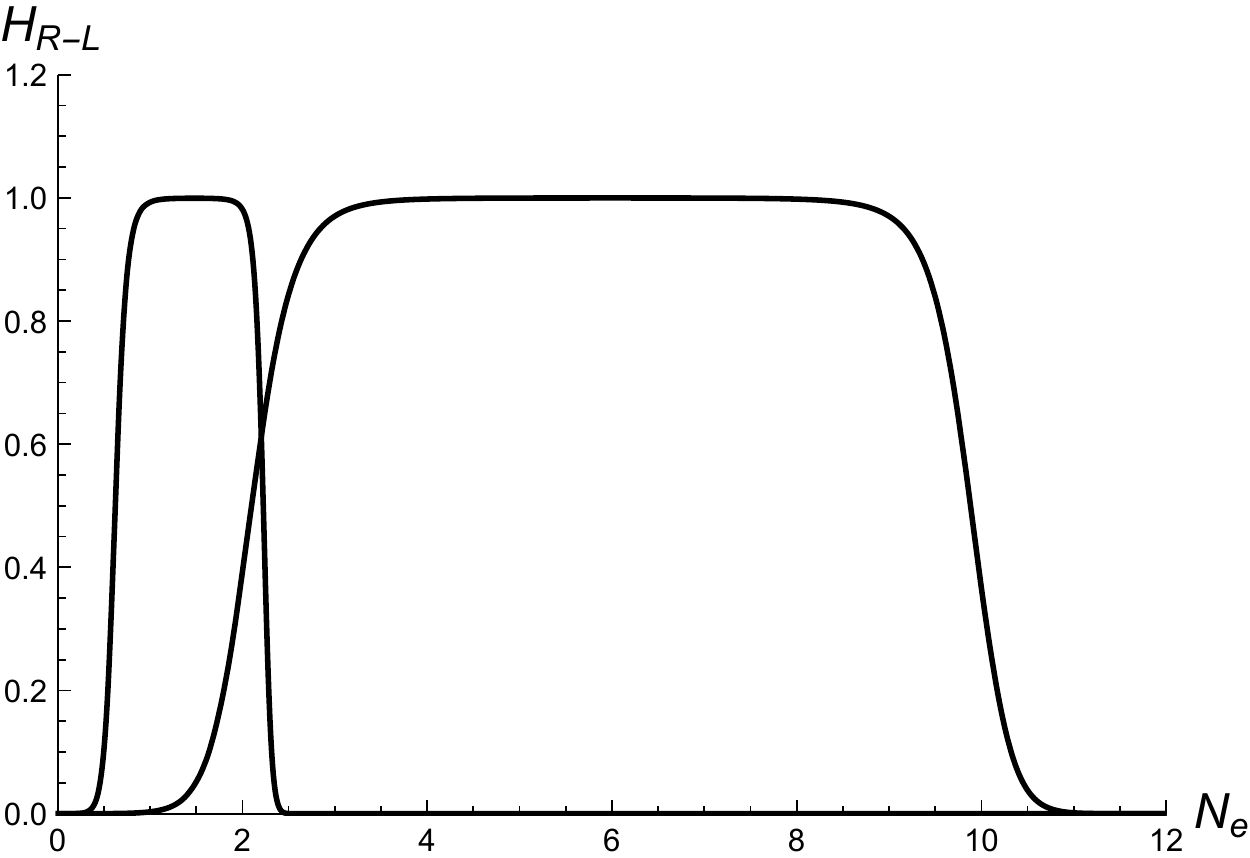}
\caption{Spectator $C_2$ with a periodic potential,  for a model with $r_{infl}=10^{-5}$ and with coupling $\lambda=10$ (left curves) and $\lambda=50$ (right curves). [{\bf Left Panel}]: Amplification of the tensor power spectrum $P_h(k)$ for modes that exit the horizon in the first 10 e-folds of inflation. The red dashed lines corresponds to a scale-invariant spectrum with $r=.01$ and $r=0.1$ respectively. [{\bf Right Panel}]: The net helicity fraction of the gravitational wave power spectrum .} 
\label{fig:AmpNPlot}
\end{figure*}

{\bf 1. \emph{Oscillatory Potential}: } Gaugino condensation on D7 branes carrying worldvolume flux generates an oscillatory potential for the $C_2$ axion $c$ of the form \cite{Long:2014dta}
\bea
&& V = \Lambda^4 \left[ 1 - \cos \left( \frac{ 2 \pi}{N} M c\right)\right] \\
&&  \Lambda^4 \simeq \frac{|W_0| A}{{\cal V}^2 N} \tau e^{- a \tau}, 
\eea
where $M$ is the number of units of flux on the branes, $\tau$ is the 4-cycle volume modulus, and $\{a,A\}$ are the parameters of the non-perturbative superpotential. We provide further details on this in section \ref{sec:ModStab}. 
\newline \newline

{\bf 2. \emph{Decay Constant}}: As shown in equation \eqref{eq:f}, the slow-roll attractor can be realized via a suitably small decay constant $f$. This naturally arises in warped flux compactifications due to the `warping down' of the decay constant \cite{Dasgupta:2008hb, Franco:2014hsa}. For example, in the Klebanov-Strassler geometry \cite{Klebanov:2000hb} the axion decay constant is given by
\be
\tilde{f} \simeq e^{- \frac{4 \pi}{3} \frac{K}{M g_s} } f ,
\ee
where $M$ and $K$ are the number of $RR$ and $NS$ flux quanta respectively, and $f$ is the decay constant in the unwarped  compactification. For $K\simeq M$ and at weak string coupling $g_s \ll 1$, this leads to a large suppression of the decay constant.

In a compactification with \emph{multiple} warped throats, the decay constant is given by \cite{Dasgupta:2008hb},
\be
\tilde{f} \simeq \frac{f}{\sqrt{\sum_i c_i ^{-2} h_i^{-2}}}  ,
\ee
where the $c_i$  are a set of  $\mathcal{O}(1)$ constants, and $h_{i}$ is the warp factor at the tip of the $i^{th}$ warped throat, which for Klebanov-Strassler is simply $h_{i} = e^{ -\frac{4 \pi}{3} \frac{K_i}{M_i g_s} }$. It follows that the axion decay constant is effectively set by the longest throat in the compactification \cite{Dasgupta:2008hb}.

This generates a large $\lambda/f$, necessary for the slow roll evolution of the axion gauge field system, while not affecting the gauge coupling of the stack of D7's, which we take to be localized far from the longest throat in the compactification. 
\newline \newline

 {\bf 3. \emph{Coupling to Gauge Fields}:} As per equation \eqref{NGW}, the number of e-folds of slow-roll evolution is determined by the size of $\lambda$, independent of $f$.

 Details of the couplings of the various string theory axions can be found in e.g.~\cite{Cicoli:2012sz}. The coupling of $C_2$ arises from Chern-Simons interaction on a D-brane carrying worldvolume flux,
\be
 \int C_2 \wedge F_2 \wedge F_2 \wedge F_ 2 = \int d^4 x \;  M c F \tilde{F} ,
\ee
where $M$ is defined as
\be
M =\frac{1}{2 \pi \alpha'} \int_{\Sigma_2} F_2 .
\ee
The coupling to the canonically normalized gauge field is $g^2 M $, where $g^2  = 1/\tau$ is the gauge coupling. When this coupling is defined as a ratio to the decay constant, $\lambda/4f$ with $f$ read off from the oscillatory potential, one finds\footnote{Carefully accounting for the $(1/4)$ factor convention \cite{Cicoli:2012sz}.}
\be
\label{lambda}
\lambda = \frac{N g^2}{2 \pi} = \frac{N }{2 \pi \tau} ,
\ee
independent of the warping down of $f$ or the number of units of flux $M$. Sizeable values of $\lambda$ can result if the $C_2$ axion wraps a small cycle in the compactification, e.g.~for $\tau=2$, one can achieve $\lambda=5$ by $N=62$.

The allowed values for $N$, and hence $\lambda$, can be probed by demanding consistency with tadpole cancellation, which in turn relates $N$ to the topology of the internal space, as discussed in \cite{Louis:2012nb}. Ref.~\cite{Louis:2012nb} found manifolds allowing for $N=\mathcal{O}(10^3)$, and with this in mind here we consider $N = \mathcal{O}(1)\times 10^2$. 

The global structure of the compactification is further constrained by demanding weak coupling to the standard model $U(1)$. which is necessary to avoid the over-production of scalar non-Gaussianities known to arise in Abelian models (see Appendix \ref{App} for a detailed discussion). This is easily achieved if the standard model is realized on a large cycle $\tau_{\rm S.M.} \gg 1$, which we will assume to be the case here.

\subsection{The Model and Gravitational Wave Signal}

Putting the puzzle pieces together, we arrive at the coupled axion gauge field system
\be
\mathcal{L} = -\frac{1}{2}\left(\partial\chi\right)^{2}-\mu^4 \cos\left( \frac{\chi}{f}\right)-\frac{1}{4}F^2+\frac{\lambda\,\chi}{4f}F\tilde{F},
\ee
with the parameters specified by: 
\bea
&&\mu^4 = \frac{|W_0| A}{{\cal V}^2 N} \tau e^{- 2\pi\tau/N} \;\; , \;\; f = \frac{\tilde{f} N }{2 \pi M} \nonumber \\
&&  \lambda = \frac{N}{2 \pi \tau} \;\; ,\;\; g^2 = \frac{1}{\tau}
\eea
where $\tilde{f}$ is the warped-down axion decay constant, and $f$ is defined as the factor appearing inside the cosine potential.

We numerically solve for the evolution of the $\{\chi,Q\}$ system in an inflationary background, and from this compute the gravitational wave signal. The amplification of the tensor power spectrum, for a host inflation model with $r_{inf}=10^{-5}$ ($H_{inf}\simeq 6 \times 10^{-7} M_{Pl}$), is given in the left panel of Figure \ref{fig:AmpNPlot}, where the two black curves correspond to $\lambda=10$ and $\lambda=50$. The other parameters are chosen as, 
\bea
&&  \mu = 8.4  \times 10^{-6} M_{Pl}, \; 1.2\times 10^{-5} M_{Pl}\;\;,\; \; \nonumber  \\
&& g^2 = \frac{1}{2}\;\; , \;\; f=10^{-12} M_{Pl}   \;\; , \;\; \chi_* = \frac{\pi f}{40} ,
\eea
where the two values of $\mu$ correspond to the $\lambda=10$ and $50$ cases respectively, and $\chi_*$ is the initial value of $\chi$. Achieving these values of $\mu$ generically requires $A \ll 1$, which we will see is also required by moduli stabilization.  

For $\lambda=10$, corresponding to $N \simeq 126$, the amplification of gravitational waves occurs for roughly an e-fold of inflation, and hence a small range of $k$ around the CMB pivot scale. On the other hand, for $\lambda=50$, corresponding to $N \simeq 628$, the amplification lasts for $\simeq 5$ e-folds.

In both cases the amplification tracks the evolution of $m_Q \propto Q \propto \sin ^{1/3} (\chi/f)$, leading to a peaked structure, with the peak corresponding to the moment when $\chi=\pi f/2$. For larger values of $\lambda$, the amplification lasts longer, scaling roughly linearly with $\lambda$, while maintaining the same peaked structure. This implies that both the tensor-to-scalar ratio $r$ and the tensor tilt $n_t$ depend sensitively on the time at which the CMB pivot scale exited the horizon, as observed previously in \cite{Fujita:2017jwq}.

\begin{figure*}[ht!]
\centering
\includegraphics[width=8.5cm]{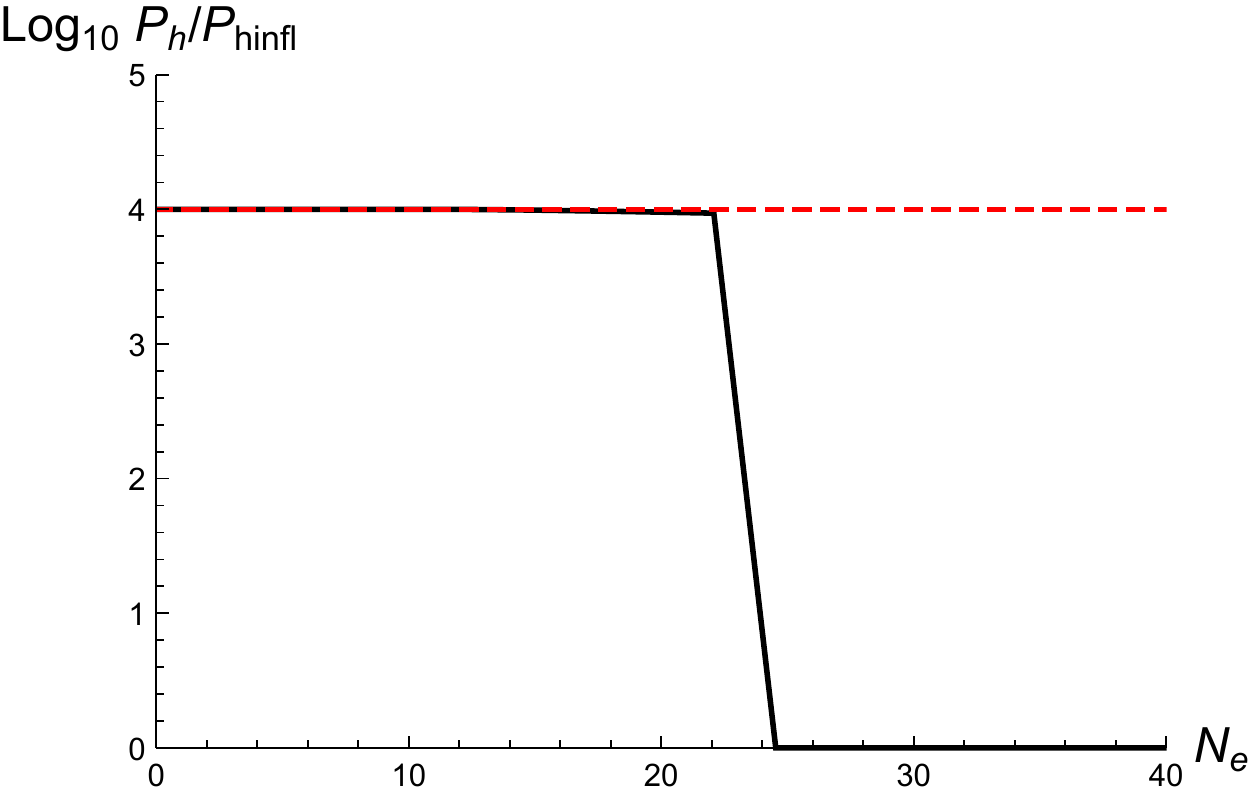}
\includegraphics[width=8.5cm]{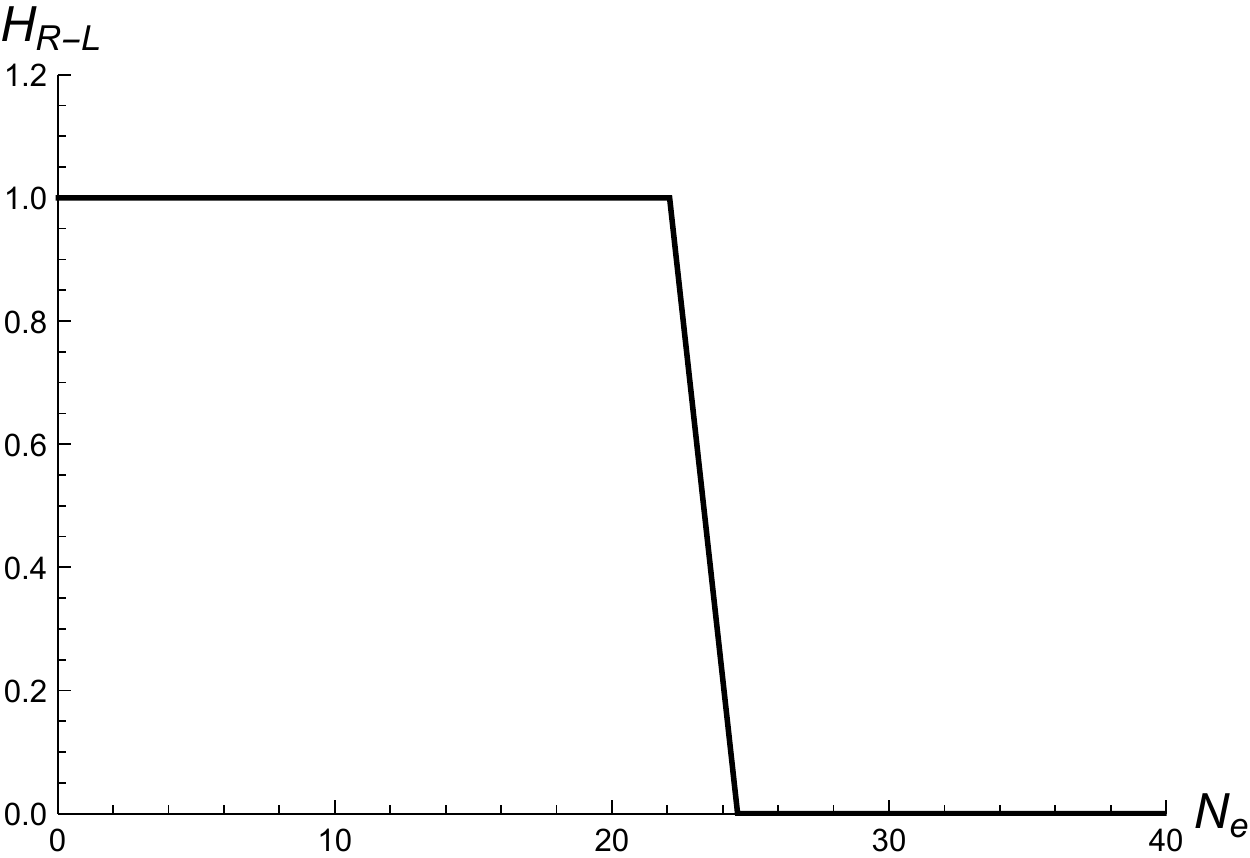}
\caption{Spectator $C_2$ with a monodromy potential,  with $r_{infl}=10^{-6}$ and with $\lambda=2$. [{\bf Left Panel}]: Amplification of the tensor power spectrum $P_h(k)$ for modes that exit the horizon in the first 40 e-folds of inflation. The red dashed line corresponds to a scale-invariant spectrum with $r=10^{-2}$. [{\bf Right Panel}]: The net helicity fraction of the gravitational wave power spectrum.}
\label{fig:MonoPlotlambda2}
\end{figure*}

The chirality of gravitational waves is also substantially impacted. One can define the helicity of the gravitational wave power spectrum as
\be
\mathcal{H}_{R-L}(k) = \frac{P_{hR}  - P_{hL} }{P_{hR}   + P_{hL} } ,
\ee
where $P_{h L,R}$ refers to the tensor power spectrum of the left,right tensor modes. The gravitational waves produced by the non-Abelian gauge fields are purely right-handed, leading to a sizeable helicity on large scales. This is shown in the right panel of Figure \ref{fig:AmpNPlot}. As discussed in \cite{Lue:1998mq, Contaldi:2008yz}, this can be probed observationally by parity violating CMB cross-correlations EB and TB.

The hallmark of this scenario is the short duration of gravitational wave production, occurring for $\mathcal{O}(1)$- $\mathcal{O}(10)$ e-folds of inflation. This is can be modified by allowing for a large hierarchy between $N$ and $\tau$, or by a \emph{monodromy} for the axion, which we now proceed to study.

%%%%%%%%%%%%%%%%%%%%%%%%
\section{$C_2$ Axion Monodromy}
\label{sec:mono}
%%%%%%%%%%%%%%%%%%%%%%%%

An alternative construction is to include a monodromy potential \cite{McAllister:2008hb,Silverstein:2008sg} for the $C_2$ axion. This monodromy potential arises e.g.~from the DBI action of an NS5 brane, leading to a potential of the form,
\be
\label{Vmono}
V_{mono} = \mu^4 \sqrt{ \left( \frac{\chi_c}{f}\right) + \left( \frac{\chi}{f}\right)^2 } ,
\ee
where $\mu$ is given by  \cite{Silverstein:2008sg}
\be
\mu^4 \equiv f \cdot \frac{\epsilon}{g_s ^2 (2 \pi)^5 {\alpha'}^2} ,
\ee
where $\epsilon$ encodes a warp factor dependence, and the value of $\chi_c$ dictates the transition from a linear to quadratic potential. As in \cite{Silverstein:2008sg}, we consider the case that $\chi_* \gg \chi_c$, where the potential is approximated as $V \simeq \mu^4 \chi/f$.

As suggested in \cite{Retolaza:2015sta}, axion monodromy can be straightforwardly accommodated into supergravity, and moduli stabilization, in the framework of F-term Axion Monodromy \cite{Marchesano:2014mla}. As a simple example, following \cite{McDonough:2016der}, one can realize this via the spontaneous symmetry breaking of a nilpotent superfield \cite{Rocek:1978nb,*Ivanov:1978mx,*Lindstrom:1979kq,*Casalbuoni:1988xh,*Komargodski:2009rz,Kallosh:2014wsa, Bergshoeff:2015jxa}, via 
\be
\delta K = \frac{1}{{\bf V }(G, \overline{G})} \cdot S \overline{S} \;\; ,\;\; \delta W = M S,
\ee
with $M$ a constant and $S$ satisfying the nilpotency constraint $S^2 = 0$. In \cite{Kallosh:2017wnt} this was dubbed `$\overline{D3}$ induced geometric inflation', but it applies more generally to brane and anti-brane constructions that spontaneously break supersymmetry \cite{Dasgupta:2016prs}.

The resulting correction to the scalar potential is decoupled from the other pieces, due to the nilpotency of $S$. The new term takes a simple form,
\be
\delta V = |D_S W|^2 = M^2 {\bf V}(G,\overline{G}) .
\ee
The monodromy potential \eqref{Vmono} is then realized by the replacement $\chi =(G - \overline{G})/\sqrt{2} i $ in \eqref{Vmono}.

The monodromy potential allows the axion to roll over a large distance in field space. This occurs provided that the monodromy potential dominates over the non-perturbative oscillatory potential, which is indeed the case when $\tau_s$ is stabilized at larger values. This qualitatively changes the gravitational wave signal, which is now amplified for a large number of e-folds, and this substantially softens the constraint on $\lambda$.

The $\{\chi ,Q\}$ system can again be numerically solved, and the resulting tensor power spectrum is shown in the left panel of  Figure \ref{fig:MonoPlotlambda2}, where we take $\lambda=2$ and assume a host inflation model with $r=10^{-6}$. The initial condition for $\chi$ is taken to be $\chi_* = - 200  f$, and we take $\mu= 1.861 \times 10^{-6} M_{Pl}$. 

This leads to an amplification of the tensor power spectrum by a factor of $10^{4}$, lifting the tensor-to-scalar ratio to $r=10^{-2}$.  The helicity fraction is shown in the right panel of Figure \ref{fig:MonoPlotlambda2}, where again we see that the gravitational wave power spectrum is maximally helical on large scales.

In contrast with the case studied in Section \ref{sec:C2}, the gravitational wave production lasts for a large number of e-folds, even for $\lambda = \mathcal{O}(1)$. Additionally, the peaked structure of the amplification is no longer present and the gravitational wave spectrum on large scales is indistinguishable from large field inflation with $r_{infl}=10^{-2}$.

\section{Realization in the Large Volume Scenario}
\label{sec:ModStab}
We now come to the consistency with moduli stabilization. Before studying the $C_2$ axion and D7-branes with worldvolume fluxes, we first study generalities of moduli stabilization and the possibility of gravitational wave production via interactions of the other axions of string theory.

\subsection{Moduli Stabilization and Axions in the Large Volume Scenario}
\label{C4}
String theory flux compactifications \cite{Dasgupta:1999ss,Giddings:2001yu} generically lead to many axions in the four-dimensional effective field theory \cite{Cicoli:2012sz}. However, not all of these, and indeed the vast majority of these, cannot realize the mechanism of \cite{Dimastrogiovanni:2016fuu}, either because they are too heavy or because their dynamics would lead to destabilization of the internal space.

In this work we focus on the large volume scenario for moduli stabilization \cite{Balasubramanian:2005zx,Conlon:2005ki}. The complex structure moduli and axio-dilaton are stabilized by the flux induced superpotential,
\be
W_0 = \int \left( F_3 - S H_3 \right) \wedge \Omega ,
\ee
where $S = 1/g_s + iC_0 $. This provides the ``universal axion'' $C_0$ with a mass that is generically of similar size to that of the dilaton, and hence stabilization of the dilaton prevents $C_0$ from having a parametrically small mass ($m^2 _{C_0}\ll H^2 \ll  m^2 _{g_s}$), though some possible exceptions to this have been studied in \cite{Blumenhagen:2014nba}. This prevents $C_0$ from slowly-rolling during inflation, and thus the $C_0$ axion is not a suitable candidate for the spectator axion of \cite{Dimastrogiovanni:2016fuu}.

We now turn to the Kahler moduli. We consider an internal manifold is of the `Swiss Cheese' type, with one large bounding 4-cycle of size $\tau_b$ and many smaller 4-cycles of size $\tau_i$. The volume is given by
\be
{\cal V} = \alpha \left(\tau_b ^{3/2} - \sum_{i=2} ^{h_{1,1}} \lambda_i \tau_i ^{3/2} \right) ,
\ee
where $\alpha$ and $\lambda_i$ are model-dependent positive constants. Moduli stabilization is achieved by a combination of $\alpha'$ corrections, appearing in the Kahler potential as \cite{Becker:2002nn}
\be
K = - 2 \log \left( {\cal V} + \frac{\hat{\xi}}{2}\right) \;\; , \;\; \hat{\xi} = -\frac{\chi \, \zeta(3)}{2 (2 \pi)^3 g_s ^{3/2}} ,
\ee 
and by introducing a non-perturbative superpotential for one or more of the small cycles. The superpotential is given by
\be
W = W_0 + \sum_{i=2} ^{h_{1,1}} A_i e^{- a_i T_i}.
\ee
The scalar potential can be expanded in a series of $1/{\cal V}$, to give
\bea
\label{potLVS}
V= && \frac{1}{\cal V }\sum_{i=2} ^{h_{1,1}}\frac{8 a_i ^2}{3 \lambda_i} A_i ^2 \sqrt{\tau_i} e^{- 2 a_i \tau_i}  \nonumber \\
&&+ \frac{1}{{\cal V}^2}\sum_{i=2} ^{h_{1,1}} 4 a_i \tau_i e^{- a_i \tau_i}  W_0 \cos (a_i \theta_i ) \nonumber \\
&&+ \frac{1}{{\cal V}^3}  \frac{3 |W_0|^2 \xi}{4} + \delta V_{up} ,
\eea
where $\delta V_{up}$ is an `uplift' potential, responsible for making the post-inflation local minimum de Sitter \footnote{The uplift term is conjectured to arise e.g.~from anti-D3 branes \cite{Kachru:2003aw} or D7 worldvolume fluxes \cite{Burgess:2003ic}, though this has been subject to considerable debate e.g.~\cite{Dasgupta:2014pma, CaboBizet:2016qsa,Bergshoeff:2015jxa,Polchinski:2015bea}. This term will not directly enter into the current analysis, which is focused on inflation, and hence we do not study it further.}.

The axions $\theta_i$ each have an oscillatory potential, which one might hope could lead to the axion-spectator behaviour necessary for gravitational waves. However, the stabilization of $\tau_i$ requires that the second term in \eqref{potLVS} be negative and hence that the $\theta_i$ be stabilized at $\cos (a_i \theta_i)=-1$ \cite{BlancoPillado:2009nw} . In contrast, in the example cosmologies considered by DFF \cite{Dimastrogiovanni:2016fuu} the axion spectator begins its cosmological evolution at $\cos(\chi/f)=0$, in which case the second term in \eqref{potLVS} vanishes and the corresponding $\tau_i$ will runaway to infinity, decompactifying the internal space. Hence, as is the case for $C_0$, none of the $\theta_i$ are suitable candidates to be the axion spectator of \cite{Dimastrogiovanni:2016fuu}.

With the $\theta_i$ stabilized at $\cos(a_i \theta_i)=-1$ moduli stabilization can proceed. Provided that one cycle is parametrically smaller then the others, $\tau_s \ll \tau_i$, such that the $\tau_s$ term dominates the superpotential, then one can analytically find stabilized solutions by varying with respect to ${\cal V}$ and $\tau_s$. The result is 
\be
\langle {\cal V} \rangle = \frac{3 \lambda_s \sqrt{\tau_s} W_0 e^{a_s \tau_s}}{4 a A} \;\;,\;\; \langle \tau_s \rangle = \left(\frac{\hat{\xi}}{2 \lambda_s} \right)^{2/3} .
\ee
The remaining moduli $\tau_{i}$ can be similarly fixed via their non-perturbative superpotentials, or else undergo inflationary dynamics.

Finally, we note that the $\tau_b$ axion $\theta_b$ also cannot realize \cite{Dimastrogiovanni:2016fuu}. The gauge kinetic function of the $C_4$ axions coupling to gauge fields is given by \cite{Cicoli:2012sz}
\be
\hat{f}_{D7} = T .
\ee
And hence the coupling of $\theta_b$ to canonically normalized gauge fields on a stack of D7 branes wrapping the $\tau_b$ cycle, rescaled by the argument of the oscillatory potential, is given by $\lambda = g^2 N/(2\pi) = N/(2 \pi \tau_b)$, as in equation \eqref{lambda}. Since $\tau_b \simeq {\cal V}^{2/3}$ is required to be large, sizeable values of $\lambda$ require extremely large values of $N$, and hence an extremely large number of condensing branes.

\subsection{Dynamics of $C_2$ axion}

\begin{figure*}[htb!]
\centering
\includegraphics[width=12cm]{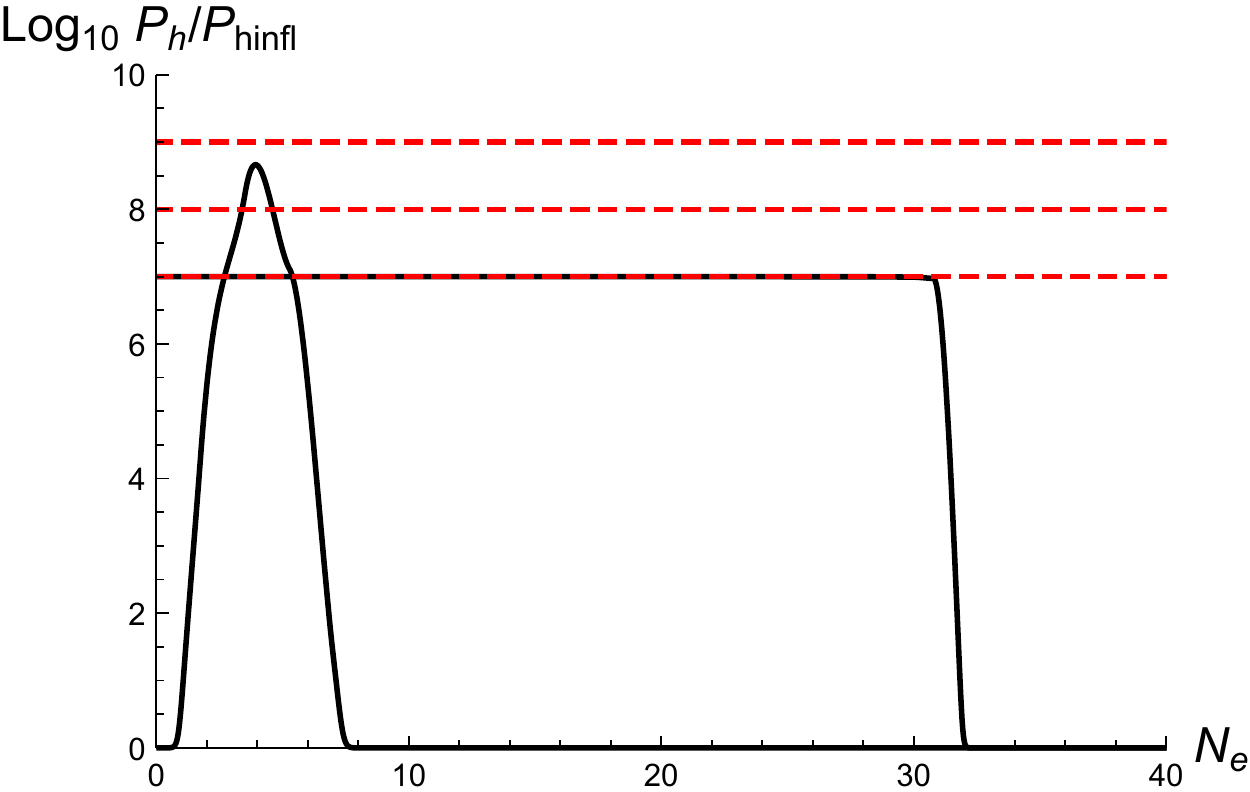}
\caption{Kahler Moduli Inflation with a spectator $C_2$. The peaked curve is an oscillatory potential with $\lambda=50$, while the flat curve is the monodromy potential with $\lambda=2$. Dashed lines indicate (from bottom to top), $r=10^{-3}, 10^{-2}$ and $10^{-1}$.  }
\label{fig:AmpNPlotKMI}
\end{figure*}

We now come to the $C_2$ axion. An important feature of the large volume scenario is that volume stabilization does not involve the $\tau_i$ and hence is decoupled from the inflationary dynamics, up to perturbative corrections. We will demonstrate that the same is true once dynamics are included for a $C_2$ axion, thus allowing for arbitrary inflationary dynamics to be added in to the model.

In the presence of worldvolume fluxes, the D7 gauge kinetic function is modified to include the two-form fields as \cite{Long:2014dta, Ben-Dayan:2014lca}\footnote{We do not consider the possibility of multiple windings, which have the effect of rescaling $\hat{f}_{D7}$ \cite{Long:2014dta}.}
\be
\hat{f}_{D7} = T + f^i G_i + \frac{1}{2 g_s} f^i f_i,
\ee
where $f^i$ are the $F_2$ flux quanta and $G_i$ contain the two-form fields,
\be
G^\alpha = \bar{S} b^{\alpha} + i c^{\alpha} ,
\ee
where again $S = e^{- \phi} + i C_0$ and
\be
c^{\alpha} = \int _{\Sigma_{2\alpha}} C_2 \;\; , \;\; b^\alpha = \int _{\Sigma_{2 \alpha}} B_2 . 
\ee
Considering only a single $C_2$ axion, the Kahler potential is given by \cite{Long:2014dta, Ben-Dayan:2014lca}
\be
K = - 2 \log \left[  \tau_b^{3/2} - \lambda_s(\tau_s + \gamma_c \tau_G ^2)^{3/2} \right] ,
\ee
where $\tau_G \equiv G + \bar{G}$, and $\gamma_c = c_{+--}/g_s$, with $c_{+--}$ the triple-intersection number of the even and odd cycles. This breaks the shift symmetry of the $B_2$ axion, while leaving that of $C_2$ intact.

The D7 world-volume fluxes introduce further subtleties, notably in the analysis of anomaly cancellation and of D-terms, which have been studied in detail in the context of D3/D7 inflation \cite{Dasgupta:2002ew}. Anomaly cancellation requires additional fluxes and brane sources, which is most easily studied in M-theory \cite{Dasgupta:2002ew}, where IIB worldvolume fluxes are described by localized contributions to the bulk $G_4$ flux \cite{Dasgupta:2008hw, Dasgupta:2014txa}.  On the IIB side, the induced D-term takes the form \cite{Jockers:2005zy, Grimm:2011dj},
\be
D = \frac{\alpha' t_{\alpha}}{2 \mathcal{V}} \kappa^\alpha _{b c} \left( b^{b} - f^b \right) {\cal W}^c,
\ee
where $t_{\alpha}$ are the two-cycle volumes, $\kappa^\alpha _{b c}$ are the triple intersection numbers, and ${\cal W}^c$ is the wrapping number of the brane. This leads to additional stabilization of $B_2$, and again leaves $C_2$ unaffected.

A potential for $C_2$ can be generated by gaugino condensation on the small cycle $\tau_s$, as studied in section \ref{sec:C2}. Here we consider that this occurs by either the condensation of a product group or else of two independent brane stacks. The superpotential is then given by a sum of two non-perturbative terms\footnote{We emphasize that both non-perturbative terms must necessarily arise from gaugino condensation, and not instantons. An analysis of the latter would require a sum over \emph{all} instanton configurations, which would generically introduce a $G$ dependence in the exponent of the first term.},
\be
\label{WLVSC2}
W = W_0 + A_s e^{- a_s T_s} + A_c e^{-a_c (T_s + M G)} ,
\ee
with $M$ the units of $F_2$ flux. Moreover, we have absorbed the $f_i f^i$ term in $\hat{f}_{D7}$ into the definition of $A_c$, and will consider the magnetized gaugino condensate to be a subdominant contribution to the superpotential, 
\be
\label{AcAs}
A_c \ll A_s.
\ee
which is the case if $f_i f^i$ is large. This could also arise if e.g. ~$A_c$ has a dependence on $\tau_b$ as $e^{- a_b \tau_b}$, or could effectively be the case if $a_c \ll a_s$.  

The scalar potential is given by, setting $a_c = a_s$ and $\gamma_c=1$ for simplicity, and with the $\theta_i$ and $B_2$ axions stabilized,
\bea
\label{potLVSC2}
V= && \frac{1}{\cal V }\frac{8 a_s ^2}{3 \lambda_s} A_s ^2 \sqrt{\tau_s} e^{- 2 a_s \tau_s}\left[ 1 + \frac{A_c}{A_s} \cos (a_c M c) \right] \nonumber \\
&& - \frac{1}{{\cal V}^2} 4 a_s A_s \tau_s e^{- a_s \tau_s}  W_0 \left[ 1 - \frac{A_c}{2A_s} \cos (a_c M c) \right] \nonumber \\
&&+ \frac{1}{{\cal V}^3}  \frac{3 |W_0|^2 \xi}{4} + \delta V_{up}  + \mathcal{O}\left(  (\frac{A_c}{A_s})^2 \right).
\eea
Comparing with \eqref{potLVS} we see that $c$-induced corrections to the scalar potential scale as $A_c/A_s$, and hence moduli stabilization of $\{\tau_s, \mathcal{V} \}$ proceeds unchanged provided that $A_c \ll A_s$. 

It follows from this that the dynamics of $C_2$ will not destabilize the internal space, nor generate any potential for the $\tau_i$. The latter implies that the $C_2$ spectator scenario is consistent with arbitrary inflationary dynamics for the $\tau_i$ moduli.

Finally, in the stabilized minimum the first two lines in \eqref{potLVSC2} are of the same order of magnitude, and the oscillatory potential for $c$ is precisely of the form anticipated in section \ref{sec:C2}.

\section{Example Host Inflation Model: Kahler Moduli Inflation}

As a concrete model example, we now consider Kahler Moduli Inflation \cite{Conlon:2005jm}. This inflation scenario arises from supplementing \eqref{WLVSC2} with a superpotential for a Kahler modulus $\tau_\phi$, of the form $A_{\phi} e^{- a_\phi T_\phi}$. The resulting model is of the DFF form \eqref{DFFL}, with inflationary potential given by
\be
V(\phi) \simeq V_0 \left[ 1 - \alpha \left(\frac{ \phi}{M_{Pl}} \right)^{4/3} e^{-  \beta (\phi/M_{Pl})^{4/3}} \right]  ,
\ee 
where $\phi \equiv \sqrt{\frac{4 \lambda_\phi}{3 \mathcal{V}} } \tau_\phi ^{3/4}$ is the canonically normalized inflaton, and the parameters are $V_0 \simeq W_0^2/{\cal V}^{3}$, $\alpha \simeq {\cal V}^{5/3}$, and $\beta \simeq {\cal V}^{2/3}$. Typical values of the volume in this scenario are $10^5 - 10^7$,  while $10^4$ can achieved by a suitably small $W_0$ \cite{BlancoPillado:2009nw}.

The precise observational predictions for Kahler Moduli Inflation have been studied by \cite{Martin:2013tda}, and are given by
\bea
&& n_s \simeq 1- \frac{2}{\Delta N_*}  , \nonumber \\
&&r \simeq \frac{4}{81 \beta^{3/2} N_* ^2} \log^{5/2} \, \left(24 \sqrt{\frac{\beta}{2}} \alpha \Delta N_* \right) ,
\eea
where $\Delta N_*$ is the number of e-folds before the end of inflation when the CMB pivot scale exited the horizon. For typical values of the parameters, this evaluates to \cite{Conlon:2005jm}
\be
0.960 \leq n_s \leq 0.967 \;\; ,\;\; r\leq 10^{-10} .
\ee
 This value of the tensor-to-scalar ratio is not obervable in the foreseeable future. The corresponding energy scale of inflation is $V_{inf} ^{1/4} \simeq 10^{13} - 10^{14}$ GeV, and the field excursion is well within the small field regime.

As an example of the amplification of $r$ due to gauge field production, we consider a model at the upper bound, $r=10^{-10}$. We consider both an oscillatory potential with
\bea
 \lambda=50 \; ,\;g^2= \frac{1}{2}  \;,\;f=10^{-12} M_{Pl}  \;\,,\;\,\mu = 5.89 \times 10^{-8} M_{Pl} \nonumber ,
\eea
corresponding to $\tau_s = 2 $ and $N=628$, and a monodromy potential with
\be
\lambda=2 \; ,\; g^2= \frac{1}{2} \; ,\; f=10^{-15} M_{Pl} \;\; ,\;\; \mu = 2.48 \times 10^{-8} M_{Pl} , \nonumber
\ee
corresponding to $\tau_s = 2 $ and $N=13$.

The resulting tensor power spectra are shown in Figure \ref{fig:AmpNPlotKMI}. As in the example in section \ref{sec:C2}, the oscillatory potential case exhibits a peaked structure, and in this example rises above that of a scale-invariant spectrum with $r=10^{-3}$ for roughly 3 e-folds of inflation. The observed tensor-to-scalar ratio $r$, and also the tensor tilt $n_t$, thus depend sensitively on the time at which the pivot scale left the horizon.  This is not the case for the monodromy potential, which exhibits an amplification to $r=10^{-3}$ for a large range of $N_e$. 

Thus we see that both spectator scenarios can realize an observably large tensor-to-scalar ratio in Kahler Moduli Inflation.

%=================================================================%
\section{Conclusion}
\label{sec:conclusion}
%=================================================================%

In this work we have studied the gravitational wave production due to excitations of non-Abelian gauge fields on D7 branes,  the presence of which is intrinsic to string inflation scenarios, due to the necessity of moduli stabilization. Worldvolume fluxes on the branes lead to an oscillatory potential for an axion, and an axionic coupling to the gauge fields. This leads to a realizations of the spectator scenario \cite{Dimastrogiovanni:2016fuu}, and is easily extended to a include a monodromy potential for the axion. 

The former case leads to a production of gravitational waves at the beginning of inflation, lasting for a few e-folds of expansion. The tensor mode power spectrum is greatly amplified during this time, and the tensor-to-scalar ratio can be lifted to an observable level, $r \simeq 10^{-3} - 10^{-2}$ \cite{Abazajian:2016yjj}.  The latter case leads to sustained production of gravitational waves, and a large amplification of $r$ even for $\lambda = \mathcal{O}(1)$.

This indicates that an observable level of gravitational waves is possible in small field inflation in string theory, once the full structure of model realizations is taken into account. Moreover, the resulting gravitational wave spectrum is maximally chiral, distinguishing it from other sources of gravitational wave production. To further quantify and constrain the observational signatures will require a full CMB polarization analysis, which we leave to future work.

We have demonstrated that this can incorporated into the Large Volume Scenario, and argued that there is negligible backreaction on moduli stabilization, or on the inflationary dynamics of the host model. However, a more detailed analysis is certainly warranted, especially given the backreaction issues known to affect monodromy models \cite{Conlon:2011qp,Hebecker:2014kva}, and recent results concerning the cosmological backreaction of gravitational waves \cite{Brandenberger:2018fte}. Related to this, in order to fully characterize the model predictions it is necessary to perform a scan of the self-consistent parameter space, both of the parameters describing the string compactification and of the those describing the cosmological model.

Finally, we mention the connection of this work to the dark matter scenario \cite{Alexander:2018fjp}. That work connected the inflationary production of chiral gravitational waves to the simultaneous generation of dark and visible particle-antiparticle asymmetries, and via the condensed matter physics of gauge theories \cite{Alford:2007xm}, to a model of superfluid dark matter. The present work is a first step in the string theory realization of that scenario.

\acknowledgments

The authors thank Keshav Dasgupta, Jerome Quintin, Ryo Namba, and Edward Wilson-Ewing, for insightful comments and suggestions. EM is supported in part by the National Science and Engineering Research Council of Canada via a PDF fellowship.

\appendix

\section{Scalar Non-Gaussianities}
\label{App}

As shown in \cite{Dimastrogiovanni:2016fuu}, the spectator sector considered here leads to negligible induced scalar perturbations, and hence no change to the power spectrum.  However, the contribution of the spectator sector to scalar non-Gaussianities requires close attention. This was recently computed in \cite{Dimastrogiovanni:2018xnn}, and confirmed to be small. Here we will put their results in context.

There are two contributions to non-Gaussianities in axion models:
\begin{enumerate}
\item Direct production of highly non-Gaussian inflaton perturbations.
\item Non-Gaussianity of the inflaton-sourced curvature perturbation induced by new interaction vertices.
\end{enumerate}
To understand these, it useful to first review non-Gaussianities in the simple case of axion inflation coupled to an Abelian gauge field \cite{Barnaby:2010vf, Barnaby:2011vw}. In that case, the dominant effect is the direct production of non-Gaussian perturbations, occuring at second-order in perturbation theory via inverse decay $A_i A^i \rightarrow \delta \varphi$ of gauge fluctuations $A_{i}$ to inflaton fluctuations $\delta \varphi$. The associated gravitational wave production is similarly given by $h_{ij} \simeq A_{i} A_j$. Since both these effects occur at second-order in $A_i$, achieving an obersvable tensor-to-scalar is incompatible with satisfying observable bounds on non-Gaussianities  .

If the axion is instead taken to be a spectator $\chi$, the inverse decay $A_i A^i \rightarrow \delta \chi$ produces entropy perturbations $\delta \chi$, whose contribution to the curvature perturbation is suppressed by $\rho_{\chi}/H$ (see e.g.~\cite{Malik:2004tf}). In this way, the effect 1.~can be neutralized.  However, effect 2.~is still present, and the coupling $\sqrt{-g} \chi F \tilde{F}$ introduces an interaction vertex \cite{Ferreira:2014zia},
\be
\label{NGint}
\mathcal{L}_{int} = \frac{\lambda \dot{\chi}}{4 f H} \zeta F \tilde{F}. 
\ee
The presence of this term can be seen by performing a time-translation $t \rightarrow t - \delta \varphi/\dot{\varphi}$ to the uniform $\varphi$ gauge ($\delta \varphi =0$), where $\varphi$ the inflaton, which simultaneously transforms $\delta\chi$ as $\delta \chi \rightarrow \delta \chi + (\dot{\chi}/H) \zeta$ \cite{Ferreira:2014zia}.

The non-Gaussianities induced by \eqref{NGint} can be computed via a Feynman diagram with a loop of gauge fields $A_i$ and three external lines of $\zeta$.  The result is that the curvature perturbation three-point function $\langle \zeta \zeta \zeta \rangle$ inherits the exponential growth of $A_i$, and is again generally in conflict with observations, unless $\lambda \ll 1$ (in which case no interesting phenomenology occurs), or else if $\dot{\chi}$ is non-zero only for a few e-folds of inflation \cite{Namba:2015gja}. 

%  The exception to this was found in , where it was shown that the model can be made consistent with observations if the 

We now come to the model studied in this paper, where a spectator axion is coupled to  \emph{non}-Abelian gauge fields in the isotropic configuration. The suppression of effect 1. is even more pronounced then in the Abelian spectator case, since the gravitational waves are produced at linear order in fluctuations $h_{ij} \simeq \langle A_i \rangle \delta A_j$, while the inverse decay is second order. The scalar sector is complicated by the presence of genuinely scalar perturbations $\langle A_i \rangle = Q \rightarrow Q + \delta Q$, but as shown in \cite{Dimastrogiovanni:2016fuu}, this system is free from instabilities provided $m_Q > \sqrt{2}$.

Effect 2.~has been calculated for the model studied here in Appendix C of \cite{Dimastrogiovanni:2018xnn}. There it was shown that demanding perturbativity of the curvature perturbation (i.e.~ that the one-loop correction to the power spectrum is subdominant to the tree-level contribution) ensures that the induced non-Gaussianity is well within observational bounds, and that both of these conditions can be satisfied for a broad range of parameter space (including that studied here).

% go over results of 1608.04216
% no scalar instability --> no production of scalars --> no contribution to scalar N-point functions. 
% In particular, sourced scalar fluctuations have amplitude \delta \phi_s /\delta \phi_vac  = 10^(-5)

\bibliography{D7_GWs}
\bibliographystyle{JHEP}

\end{document}